\documentclass[pre,twocolumn,showpacs]{revtex4}

\usepackage{epsfig}

\begin{document}

%\titleTime transformation for random walks in the quenched trap model}
\title{Weak subordination breaking for the quenched trap model}
\author{S. Burov, E. Barkai }
\affiliation{ Department of Physics, Institute of Nanotechnology and Advanced Materials,  Bar Ilan University, Ramat-Gan
52900, Israel}

%%%%%%%%%%%%%%%%%%%%%%%%%%%%%%%%%%%%%%%%%%%%%%%%%%%%%%%%%%%%%%%%%%%%%%%%%%%%%%%
%
% A B S T R A C T 
%
%%%%%%%%%%%%%%%%%%%%%%%%%%%%%%%%%%%%%%%%%%%%%%%%%%%%%%%%%%%%%%%%%%%%%%%%%%%%%%%
\begin{abstract}
 We map the problem of diffusion in the quenched trap model
onto a new stochastic process:  
Brownian motion which is terminated at the coverage ``time"   
${\cal S}_\alpha =\sum_{x=-\infty} ^\infty  (n_x)^\alpha$ 
 with $n_x$ being the number of visits to 
site $x$.  
Here $0<\alpha=T/T_g<1$ is a measure of the disorder in
the original model. This mapping allows us to treat the 
intricate correlations in the underlying random walk in the random
 environment.
The operational  ``time" ${\cal S}_\alpha$ is
changed to laboratory time $t$ with a L\'evy time  transformation.
 Investigation 
of Brownian motion stopped at ``time" ${\cal S}_\alpha$ yields 
the diffusion front of the quenched trap model which is
favorably compared with numerical simulations.  In the zero temperature 
limit of $\alpha\to 0$
we recover the renormalization group solution obtained by C. Monthus. 
Our theory surmounts critical slowing down which is found when $\alpha \to 1$. 
Above the critical dimension  two 
mapping the problem to a continuous time random walk
becomes feasible though still not trivial.

\end{abstract}

% 05.40.Jc Brownian motion
% 05.20.-y classical statistical mechanics
% 02.50.-r Probability theory stochastic processes, and statistics
% 46.65.+g Random phenomena and media

\maketitle

% suggested titles:
%{\bf Weak sub-ordination breaking for
%the diffusion front in the quenched trap model}

%{\bf
% Diffusion front in the quenched trap model}
% conundrum- an intricate and difficult problem

\section{Introduction}

 Random walks in disordered systems with a diverging expected waiting time 
have attracted vast interest over many decades. Two 
approaches
in this field are the annealed continuous time random walk (CTRW) model
and 
the quenched trap model (QTM).  Starting in the 70's,
the Scher-Montroll CTRW  approach was  used
to model sub-diffusive  photo-currents in amorphous materials 
\cite{Scher,Shlesinger,Shles,Silbey} and for contaminants transport
in hydrology \cite{Berk}. 
  Bouchaud showed that
the trap model
is a useful tool for the description of aging phenomena
in glasses \cite{B92,MB96,Biroli10}. Then fractional kinetic equations which
 describe
CTRW dynamics became a popular tool \cite{Review}. 
 More recently
these models were used to describe non self averaging \cite{Stas,Akimoto}
and weak ergodicity breaking \cite{B92,Golan} which
is important for the statistical  description 
of dynamics of single quantum dots \cite{PhysToday}   
and single molecules in living cells \cite{Yong,Soko}. 

 This manuscript presents a new approach for random walks in a quenched
 random
environment i.e. site disorder at each lattice point is
fixed in time.  
 In its generality
this topic has attracted tremendous interest in Physics
\cite{Review,Alexander,Bouchaud,comtet,Hughes,Fisher,Havlin,BenNaim} 
and Mathematics \cite{Sinai,Golosov,Kesten}. 
For  the QTM
the critical dimension 
is  two
\cite{Bouchaud,Machta,Alexander1,Math,BenHarosh}. 
Above two dimensions the Scher-Montroll
continuous time random walk (CTRW), which is a mean field theory,
qualitatively describes the sub-diffusive process. According to Polya's
theorem \cite{Weiss,Redner}  on a simple lattice and
in dimension three, a random
walk is non recurrent. Hence in a disordered system the particle
(roughly speaking) tends to visit new lattice points along its path.
In contrast in one dimension the random walk is recurrent, and a particle
visits the same lattice point many times. Thus above the critical
dimension the CTRW approach works well, but fails  in one dimension,
due to correlations of the random walk with the disorder.
In other words
renewal theory used within the annealed CTRW framework 
is not a valid description of the
QTM \cite{Bouchaud}. 
Beyond mean field renormalization group methods are used to tackle
the problem of random walks in quenched environments 
\cite{Machta,Fisher,Monthus,Dual}. 
For example Machta \cite{Machta} found the 
scaling exponents of the QTM
and Monthus \cite{Monthus} investigated
 its diffusion front in the limit of zero temperature (see details below).
While the renormalization group method is powerful,
it  has its limitations: 
a simple
approach which can predict the  diffusion front of random walkers
in the QTM
is still missing. 

 We provide a new approach for random walks in the QTM which we
call weak subordination breaking. 
 For CTRW 
it is well known that one may decompose
the process into ordinary Brownian motion and a L\'evy time process,
 an approach
called subordination  \cite{Saichev,BarkaiPRE,Meerschaert,Sokolov,Yutse}.
 In this scheme
normal  Brownian motion takes place in an operational time $s$.
The disorder is effectively described by a L\'evy time transformation from
operational time $s$ to  laboratory time $t$ (see details below). 
This method is not intended for random walks in fixed random environments
since it is based on the renewal assumption. The latter implies the
neglect of correlations in the sense that waiting times are
not specific to a lattice site. 
So a new approach
capable of dealing with quenched
disorder is now investigated.  
A brief summary of our results was published in \cite{BurovPRL}. 

 This manuscript is organized as follows. After presenting the 
QTM in Sec. 
(\ref{SecQTM}) we briefly review the standard subordination scheme 
in Sec. 
(\ref{SecSUB}). The concept of random time in the QTM 
is presented in \ref{SecTime} which leads to   weak sub-ordination breaking 
in Sec \ref{SecWSB}. 
General properties of the diffusion front $\langle P(x,t)\rangle$ are
found in Sec. \ref{SecDIF} while Sec. 
\ref{Secalz} and \ref{Secalo} deal with the limits strong  and
weak disorder respectively. 
Sec. \ref{secSim} discusses critical slowing down. All along the
work we compare theory with numerical simulations. 

\section{Quenched Trap Model \cite{Bouchaud,BenHarosh,Bertin,linRes}} 
\label{SecQTM}

 We consider a random walk on a one dimensional lattice with lattice spacing
equal one. For each lattice
site  $x$ there is a quenched random variable $\tau_x$ which is the 
waiting time between jump events for a particle situated on $x$. After time
$\tau_x$ has elapsed 
the particle jumps to one of its two nearest 
neighbors with equal probability. 
The particle starts on the origin $x=0$ at time $t=0$, waits for time
$\tau_0$, then jumps (with probability $1/2$) to $x=1$, waits there for
$\tau_1$ etc. Note that if the particle returns to $x=0$ it will wait there
again for a time interval $\tau_0$. 
The $\{\tau_x\}$s are positive independent identically
distributed random variables with a common PDF $\psi(\tau_x)$.  
The goal of this paper is to find the long time behavior
of $\langle P(x,t) \rangle$,
the probability of finding the particle on $x$ at time $t$
averaged over the disorder. 

 In the literature two related models are usually considered. 
The first model, which we use in simulations presented below,
assumes that for a given lattice site $x$ a particle will wait
for a fixed waiting time $\tau_x$. A slightly more physical approach is to
assume that waiting times on lattice point $x$ are exponentially
distributed with a mean $\tau_x$. 
Bertin and Bouchaud \cite{Bertin},
showed that the two approaches yield the same asymptotic results
in the limit of long measurement times. 
 
 In this manuscript our main interest is with power law waiting times
\begin{equation}
\psi(\tau_x) \sim { A \over |\Gamma(-\alpha)| } (\tau_x)^{- (1 + \alpha)}  
\label{eqTR01}
\end{equation} 
for $\tau \to \infty$ and $0<\alpha<1$. The mean waiting
time  $\langle \tau_x\rangle=\infty$ and in this sense the diffusion
is scale free. 
According to Tauberian theorem  \cite{Weiss} 
the Laplace transform of the waiting time PDF  is
\begin{equation}
\hat{\psi}(u)\sim 1 - A u^\alpha + \cdots 
\label{eqTR02}
\end{equation} 
when $u \to 0$. 
In the QTM the physical mechanism
leading to these power laws is based on trapping dynamics \cite{Bouchaud}. 
On a lattice points $x$ we randomly assign traps. The energy depth of the
trap on $x$ is $E_x$ and the process of activation from a trap is thermal.
According to  Arrhenius law
 $\tau_x \propto \exp(E_x/T)$ where $T$ is the temperature.
Then assume that the 
 PDF of $E_x>0$ is exponential $f(E_x)= \exp( - E_x/T_g) /T_g$ where
$T_g$ is a measure of the energy disorder. One easily finds
\begin{equation} 
\alpha={T\over T_g} \ \ \mbox{and} \ \ 
A= |\Gamma(-\alpha)| \alpha.
\label{eqTR02aa}
\end{equation} 
Due to the Boltzmann factor $\tau_x \propto \exp(E/T)$ small
changes in energy lead to exponential changes in waiting times,
thus it is enough to have an exponential distribution of energy
traps to obtain power law waiting times. 
Experimental observation of the linear dependence of $\alpha$ on
temperature, in photocurrent spectroscopy  in a -$\mbox{As}_2 \mbox{Se}_3$
can be found in Fig. 3 in \cite{Oren}.
We note that the stochastic dynamics under investigation
 describes several other
mechanisms of anomalous diffusion beyond the QTM \cite{Stas}.
For  example random walks  on comb structures with
power law distributed
lengths of the comb's teeth which mimics a random walk
on the percolation
cluster. Thus the trap model describes
both energetic disorder and spatial disorder.

In what follows we will also consider the limit $\alpha \to 1$.
This limit is meant in the sense that $\hat{\psi}(u) \sim 1 - A u ....$ 
which means that the average waiting time is finite (Gaussian diffusion front).
The  very special border case $\psi(\tau_x) \propto \tau^{-2}$ 
was treated by Bertin and
Bouchaud  \cite{Bertin}. 
It yields Gaussian diffusion with logarithmic corrections
and is not treated here. 

\section{Subordination in the annealed trap model (=CTRW)}
\label{SecSUB}

 We now briefly review the annealed version of the model:
the well investigated Scher-Montroll-Weiss continuous time random
walk (CTRW) \cite{Bouchaud,Review,Weiss}
in particular we discuss the concept of time subordination 
\cite{Saichev,BarkaiPRE,Meerschaert,Sokolov,Yutse}. 
Later  we contrast the CTRW approach  with the intricate problem of
the quenched type.  The CTRW model considered here is 
for a one dimensional random walk on a lattice with
lattice spacing equal unity. Starting on the origin $x=0$
at time $t=0$  the particle
waits for a time $t_1$ 
it then jumps to one
of its nearest neighbors (lattice points $x=+1$ or $x=-1$) 
with equal probability.
The process is then renewed, namely the
particle waits on lattice point $+1$ (for example) for time $t_2$
 until it jumps back to $0$ or $+2$ etc. The waiting 
times $\{t_1,t_2, \cdots,t_n,\cdots\}$
are independent, identically distributed random variables
with a common PDF $\phi_{\alpha}(t)$. Here
$t_n$ is the $n$th waiting time, which is not correlated
with a specific lattice point $x$ and hence clearly the CTRW model
is very different from the quenched case. Similar to the quenched case
we consider waiting time PDFs with a diverging averaged waiting times
$\int_0 ^\infty t \phi_{\alpha}(t) {\rm d} t =\infty$ namely
\begin{equation} 
\phi_\alpha\left(t\right) \sim{ A_{\alpha} \over | \Gamma(-\alpha)|} t^{ - (1 + \alpha)}, 
\label{eqadd0}
\end{equation} 
with $0<\alpha<1$ and $A_{\alpha}>0$. 
The corresponding  Laplace transform of the waiting time PDF
behaves like
\begin{equation}
\hat{\phi_{\alpha}}(u) \sim 1 - A_{\alpha} u^\alpha + \cdots 
\label{eqCTRW0} 
\end{equation} 
when $u$ is small. 
As well known the diffusion 
 is anomalous 
$\langle x^2 \rangle \propto t^\alpha$
\cite{Bouchaud,Review}. 

 Let $P(x,t)$ be the probability of finding the particle on $x$ at time
$t$. By conditioning on the number of jumps $s$ performed
till time $t$  
\begin{equation} 
P(x,t) = \sum_{s=0} ^\infty n_t(s) q(x,s) 
\label{eqCTRW1} 
\end{equation}
where $n_t(s)$ is the probability of performing $s$ jumps in
time interval $(0,t)$ and $q(x,s)$ is the probability that after $s$
steps the particle is located on $x$. In the limit of large $t$
the number of jumps $s$ is also large. Following \cite{Bouchaud} 
 we apply the Gaussian central limit theorem
\begin{equation}
q(x,s) \sim {1 \over \sqrt{2 \pi s} } e^{ - {x^2 \over 2 s}}  
\label{eqCTRW2}
\end{equation} 
which is valid when $s \to \infty$.
Here we used the model assumption that
the variance of the jump lengths is unity
i.e. the lattice spacing is equal $1$. 

 To find $n_t(s)$ we  consider the random time
\begin{equation} 
t=\sum_{i=1} ^s t_i. 
\label{eqCTRW2a} 
\end{equation} 
In the limit of large $s$
the time is  a sum of many  
independent identically distributed random variables with a diverging
mean waiting time (since $0<\alpha<1$).  Hence L\'evy's limit theorem 
applies.
Let  
\begin{equation} 
\eta={ t \over s^{1/\alpha} }
\label{eqCTRW2b} 
\end{equation} 
then in the $s\to \infty$ limit
\begin{equation} 
\langle e^{ - u \eta} \rangle= \hat{\phi_{\alpha}}^s \left( { u \over s^{1/\alpha} }\right)= \left( 1 - { A_{\alpha} u^\alpha \over s} + \cdots \right)^s \to
 e^{ - A_{\alpha} u^\alpha} .
\label{eqCTRW3} 
\end{equation} 
Namely the PDF of $\eta>0$ is a one sided
L\'evy function denoted with $l_{\alpha,A_{\alpha},1} (\eta) $ which is defined
via its Laplace pair
\begin{equation} 
\int_0 ^\infty e^{ - u \eta} l_{\alpha,A_{\alpha}, 1} (\eta) {\rm d} \eta = e^{ - A_{\alpha} u^\alpha} . 
\label{eqCTRW4}
\end{equation} 
These L\'evy PDFs  are well investigated:
their series expansion, asymptotic behaviors, and graphical presentations
can be found in \cite{Schneider1,Taqqu,Penson}.
 Information on these PDFs essential
for our work are summarized in Appendix A.  
From the PDF of $\eta$ we find the PDF of $s$. Since both 
$s$ and $t$ are increasing along the process the transformation is straight
forward. Using Eq. (\ref{eqCTRW3}) and 
$\eta^{-\alpha}= s/t^{\alpha}$
we find
the well known  PDF of $s$ \cite{Bouchaud} 
\begin{equation}
n_t(s) \sim { t \over \alpha} s^{-{1 \over \alpha} - 1} l_{\alpha, A_{\alpha}, 1} \left( { t \over s^{1/\alpha} } \right). 
\label{eqCTRW5}
\end{equation} 
In the long time limit the Green function of the CTRW process
is thus given by \cite{Bouchaud,Saichev} 
\begin{equation}
P(x,t) \sim \int_0 ^\infty n_t(s) { e^{ - { x^2 \over 2 s} } \over \sqrt{ 2 \pi s} }{\rm d} s
\label{eqCTRW6}
\end{equation} 
where we switched from a summation in Eq. 
(\ref{eqCTRW1}) 
to integration.

 The time transformation Eq. (\ref{eqCTRW6}) maps normal Gaussian
diffusion to  
anomalous diffusion.   
In \cite{BarkaiPRE}  $P(x,t)$  was obtained in $d$ dimension  by solving 
the integral transformation which applies more generally to 
solutions of the fractional time Fokker-Planck equation \cite{MBK}.  
More importantly,  we may think
of $s$ as an operational 
 time in which usual Brownian motion is performed. The operational time
$s$ is 
a random variable whose statistics is determined by the PDF
$n_t(s)$ where $t$ is a laboratory time.  
In other words the annealed disorder turns the operational time
to a random variable. 
We note that subordination scheme can be formulated
for the trajectories of the corresponding paths, in a continuum limit
of the walk and was the topic of extensive research
\cite{Fog,Han,Weron,Marcin,West,Friedrich,Dybiec,Henry}.

Not surprisingly 
 subordination of this type does not work for the QTM
in one dimension.  As mentioned in the introduction
the process of a random walk in
a QTM is clearly not a simple renewal process.
The particle returning to a lattice point visited already, 
``remembers" its waiting time there. 
Mathematicians have rigorously shown that 
in dimensions higher than one \cite{Math}
or in the presence of a bias \cite{Zindy} 
(see also \cite{Dentz,Kantz,Khoury}) the
CTRW approach describes  well the quenched dynamics since the
 particle does not
tend  to revisit the same lattice points many times, thus confirming
physical insight in \cite{Bouchaud,Alexander,Machta}
(for dimension $d=2$ logarithmic corrections are also important).
While the three dimensional QTM belongs to the domain
of attraction of the CTRW the calculation of the anomalous diffusion
constant is not trivial (see discussion in the summary).
Here we focus
our attention on the unsolved case: the QTM in one
dimension since there the Scher-Montroll
CTRW picture 
\cite{Weiss,Review,Bouchaud}
breaks down. 

\section{Time in the quenched trap model}
\label{SecTime}

The time $t$ in the QTM is 
\begin{equation}
t = \sum_{x=-\infty} ^\infty n_x \tau_x 
\label{eqtime01}
\end{equation}
where $n_x$ is  
 the number of visits to lattice point $x$
which we call the visitation number of site $x$. 
Since we are
interested in $\langle P(x,t) \rangle$ where the brackets
are for an average over the disorder,   
we will consider ensembles of paths on a large
ensemble of realizations of disorder. 
As mentioned the  $\{ \tau_x \}$s  are independent identically
distributed random variables with a common PDF $\psi(\tau_x)$ and
the  $\{n_x\}$s are also random variables. 

 Let us consider the random variable 
\begin{equation}
\eta = {t \over ({\cal S}_\alpha)^{1/\alpha} } 
\label{eqtime02}
\end{equation}
where
\begin{equation}
{\cal S}_\alpha = \sum _{x=-\infty} ^\infty (n_x)^\alpha
\label{eqtime03}
\end{equation}
and we call ${\cal S}_\alpha$ the $\alpha$ coverage time. 
At this stage it is convenient to consider paths where ${\cal S}_\alpha$ 
is fixed and $t$ is random and later we will switch to the opposite situation
(similar to the arguments for $s$ and $t$ in the CTRW model). 
When $\alpha=1$,  ${\cal S}_{\alpha}$ is the total number of jumps
made $\sum_{x=-\infty} ^\infty n_x = s$.
In the opposite
limit $\alpha\to 0$, 
the $\alpha$ coverage time 
${\cal S}_0$ is the  distinct number of  sites
visited by the random walker which is called the span of the random
walk. 
Notice that $t$ in Eq. 
(\ref{eqtime01}) is a sum  of non independent 
and non identical random variables.

 We show that the PDF of 
$\eta$, in the limit ${\cal S}_\alpha \to \infty$ 
is a one sided L\'evy stable function
\begin{equation}
\mbox{ The PDF of \ }  \eta \mbox{ \ is:\ } l_{\alpha,A,1} 
\left( \eta\right).  
\label{eqtime04}
\end{equation} 
Namely the heavy tailed distribution of the waiting times ${\tau_x}$ 
determines the statistics of $\eta$ through the characteristic
exponent $\alpha$, while the visitation numbers $\{ n_x \}$ determine
the scaling through ${\cal S}_\alpha$.  
By definition
the Laplace $\eta \to u$ transform of the PDF of $\eta$ is 
\begin{equation}
\langle e^{ - \eta u } \rangle = \langle \exp\left[  - \sum_{i=-\infty} ^\infty {  n_i \tau_i \over \left( {\cal S}_\alpha \right)^{1/\alpha} } u \right] \rangle.
\label{eqApA01z}
\end{equation} 
We average with respect to the 
disorder, namely with respect to the independent and identically
distributed  random waiting times $\tau_x$,
and obtain
\begin{equation}
\langle e^{- u \eta } \rangle = \Pi_{x=-\infty} ^\infty \hat{\psi} \left[ { n_x u \over  ({\cal S}_\alpha)^{1/\alpha} }  \right] 
\label{eqApA02z}
\end{equation} 
where $\hat{\psi}(u)$ is the Laplace transform of the PDF of waiting
times $\psi(\tau_x)$. Now assume
 $\hat{\psi}(u) = \exp( - A u^\alpha)\sim 1 - A u^\alpha+ \cdots$.
Then using
Eq. (\ref{eqApA02z}) we have 
\begin{equation} 
\langle e^{ - u \eta} \rangle = \Pi_{x=-\infty} ^\infty \exp\left[ - {A (n_x)^\alpha u ^\alpha \over {\cal S}_\alpha }\right] = e^{ -A u^\alpha}. 
\label{eqApA02ez}   
\end{equation}   
Hence if the waiting PDF is a one sided L\'evy PDF, i.e. 
$\hat{\psi}(u) =\exp(- A u^\alpha)$,  so is the PDF
of $\eta$. In Appendix B we consider the general
case  
where $\psi(\tau_x)$ belongs to the
domain of attraction L\'evy PDFs 
(i.e. families  of PDFs satisfying Eq. $\hat{\psi}(u) \sim 1 - A u^\alpha + \cdots$).
We there prove that the 
statement in Eq.
(\ref{eqtime04}) is valid. 

 From Eq. (\ref{eqtime04}) we learn that
 the CTRW operational time $s$, that is the number
of jumps made in the random walk, looses its importance in the quenched model.
{\em In the QTM  the operational time is} the $\alpha$ coverage time
 ${\cal S}_\alpha$. 
We now invert the process fixing time $t$ to find
the PDF of ${\cal S}_\alpha$ 
\begin{equation}
n_t \left( {\cal S}_\alpha \right) \sim {t \over \alpha}  ({\cal S}_\alpha)^{ - 1/\alpha - 1} l_{\alpha,A,1} \left[ { t \over ({\cal S}_\alpha )^{1/\alpha} }\right]. 
\label{eqtime05}
\end{equation}
In the next section we explain how to use
the operational time ${\cal S}_\alpha$ to obtain the
desired diffusion front of the QTM.

\section{Weak Subordination Breaking} 
\label{SecWSB}

 To find the solution of the problem, namely find
$\langle P(x,t)\rangle$ for the QTM
we follow the following steps:
\begin{itemize}
\item[1.] Choose the laboratory time $t$ which is a fixed parameter.
\item[2.] Use a random number generator and draw the stable
random variable  $\eta$ from the
one sided L\'evy law $l_{\alpha,A,1} (\eta)$.
\item[3.] With $\eta$ and $t$ determine the hitting target
${\cal S}_\alpha$ which  
according to Eq. (\ref{eqtime02}) is
${\cal S}_\alpha= (t/\eta)^\alpha$.
\item[4.] Generate  a Binomial random walk on a lattice, with probability
$1/2$ for jumping left and right. Stop the process once its 
${\cal S}_\alpha$ crosses the hitting target set in step $3$.   
\item[5.] Record the position $x$ of the particle at the end of the previous
step.
\item[6.] Go to step 2. 
After this loop is repeated many times,
 we generate a histogram of $x$. 
\end{itemize} 
 The histogram  once normalized  
yields $\langle P(x,t) \rangle$ when $t$ is large. 
Notice that in this scheme there is no disorder. 
The second step is implemented with a simple
algorithm provided by Chambers et al \cite{Chambers}.
These authors  show how  to generate stable
random variables like $\eta$ using two independent uniformly
distributed random variables and for convenience their formula is  
provided in Appendix A. 

More importantly we can now
start treating the problem analytically, and find the diffusion front. 
So far we have replaced the problem
of random walks in the QTM, with a new stochastic process: Brownian motion
which is stopped when the hitting target ${\cal S}_\alpha$ is crossed.
In other words we got rid of the disorder. 
Notice that  the CTRW process and standard 
subordination \cite{Saichev,BarkaiPRE,Meerschaert,Sokolov,Yutse}
are reached once we replace ${\cal S}_\alpha$ with
$s$.   
In this sense the QTM exhibits what we call weak 
sub-ordination
breaking:  the operational time is now ${\cal S}_\alpha$ 
still 
the L\'evy time transformation used already in the usual subordination
scheme 
Eq. (\ref{eqCTRW6}) 
remains a useful tool.

\section{The diffusion front of the quenched trap model}
\label{SecDIF}

 Let $P_B(x,{\cal S}_\alpha)$ be the PDF of $x$ for a binomial
random walk on a lattice at operational time
${\cal S}_\alpha$. The subscript $B$ indicates that the underlying motion
is Brownian.
 The corresponding  paths are generated   from a random
walk on a one dimensional lattice,
 with equal probability of jumping left and right,
which is stopped when ${\cal S}_\alpha$ is reached (or crossed for the first
time). The averaged over disorder  propagator of the QTM 
is found using Eq. (\ref{eqtime05}) and the scheme presented in the last
section:
\begin{equation}
\langle P(x,t) \rangle \sim \int_0 ^\infty P_B(x,{\cal S}_\alpha)  
n_{t} \left( {\cal S}_\alpha\right) {\rm d} {\cal S}_\alpha, 
\label{eqP01} 
\end{equation} 
which is valid in the  long time  limit.
Eq. (\ref{eqP01}) is a generalization of the subordination
equation (\ref{eqCTRW6}). Namely it
transforms Brownian motion stopped at the coverage
time ${\cal S}_\alpha$ to the QTM dynamics in laboratory time $t$.

From Eq. (\ref{eqP01}) we may   find general properties
of the Green function $\langle P(x,t) \rangle$ in terms of
its corresponding Brownian partner $P_B(x,{\cal S}_\alpha)$. For
example the Laplace $t \to u$ transform 
\begin{equation}
\langle \hat{P} (x, u ) \rangle = A u^{\alpha -1} \hat{P}_B (x, A u^\alpha).  
\label{eqP02} 
\end{equation} 
Less formal relations are found if we exploit the scaling behavior
of Brownian motion as now explained. 

\begin{figure}
\begin{center}
\epsfig{figure=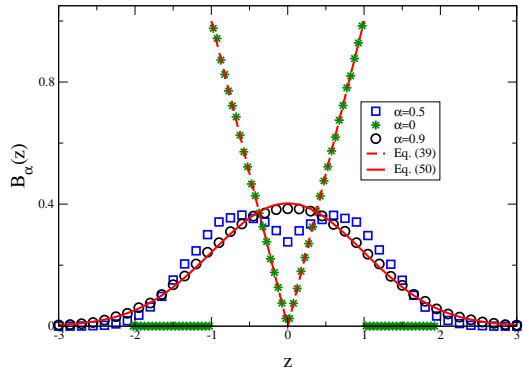,totalheight=0.34\textheight,angle=-90,width=0.45\textwidth}
\end{center}
\caption{ The PDF  $B_\alpha (z)$ exhibits a transition between a Gaussian
shape when $\alpha \to 1$ to a $V$ shape when $\alpha\to 0$.  
Simulations of Brownian motion on a lattice yield excellent agreement
with theoretical predictions Eqs.  
(\ref{eqaplz05},
\ref{eqBa102a}) without fitting.  
}
\label{fig1}
\end{figure}

\subsection{Scaling arguments} 

 Brownian  motion
follows the  usual diffusive scaling $x^2 \propto s$ where $s$ is the number
of steps. In Appendix C we 
show that $s \propto ({\cal S}_\alpha) ^{ 2/(1 +\alpha)}$
 which is now explained using simple  arguments.
For Brownian motion the particle explores a region which
scales like $s^{1/2}$. The visitation
number $n_x$
 within this region (i.e. roughly  $|x|<s^{1/2}$)  is 
the number of jumps made (s) divided by the number of sites 
in the explored
region $(s^{1/2})$ so
$n_x \propto 
s/s^{1/2}= s^{1/2}$.  Since particles typically visit
$|x|>>s^{1/2}$ only rarely $n_x\propto 0$ there.  
Hence ${\cal S}_\alpha \propto \sqrt{s} (n_x)^\alpha  \propto s^{(1+\alpha)/2}$.
Indeed in Appendix C we show that 
\begin{equation}
\langle {\cal S}_\alpha \rangle = {\cal C}_\alpha s^{{ 1 + \alpha \over 2} } 
\label{eqApC1a3xx} 
\end{equation} 
with
\begin{equation}
{\cal C}_\alpha = { 2^{{\alpha + 3 \over 2} } \Gamma\left( 1 + {\alpha \over 2}\right) \over \sqrt{\pi} \left( 1 + \alpha \right) } .
\label{eqApC14zz} 
\end{equation} 
When $\alpha=1$ we have ${\cal C}_1=1$ since
 ${\cal S}_1 = \sum_{x=-\infty} ^\infty n_x = s$. In the opposite limit
$\alpha=0$ we find  a well known  result
  obtained by Dvoretzky and Erd$\ddot{\mbox{o}}$s \cite{Erdos} 
\begin{equation}
\langle {\cal S}_0 \rangle =  \sqrt{ { 8 s \over \pi}}   .
\label{eqApC14azz} 
\end{equation} 
By definition $\langle {\cal S}_0 \rangle$ is 
the averaged number of distinct sites visited by an unbiased random
walker \cite{Weiss}.

\begin{table}[ht]
\centering
\begin{tabular}{ c c c}
\hline\hline
$\alpha$ & $\langle z^2 \rangle$ & $B_\alpha (z=0)$ \\  [0.5ex]
\hline

\ 0 \ & \ 0.5 \  &\  0 \ \\
 \ 0.1 \  & \  \  0.592 \  & \ 0.08 \ \\
 \ 0.2 \  & \  0.673 \  & \ 0.15 \  \\
 \ 0.3 \  & \  0.746 \  & \ 0.2 \  \\
 \ 0.4 \  & \  0.808 \  & \ 0.24 \  \\
 \ 0.5 \  & \  0.859  \ & \ 0.28 \  \\
 \ 0.6 \  & \  0.907 \  & \ 0.3  \ \\
 \ 0.7 \  & \  0.929  \ & \ 0.33 \  \\
 \ 0.8 \  & \  0.961  \ & \ 0.35 \  \\
 \ 0.9 \  & \  0.986 \  & \ 0.38 \  \\
 \ 1 \ & \ 1 \   &  \ $1/\sqrt{2\pi}$ \  \\

\hline

\end{tabular}
\caption{Brownian simulations on a lattice give $\langle z^2 \rangle$ and
$B_\alpha(z=0)$ which in turn provide the corresponding solution of
the QTM with Eqs.  
(\ref{eqSca07},
\ref{eqSca08}).} 
\label{table1}

\end{table}

 Using $x\propto s^{1/2}$  and 
${\cal S}_\alpha \propto s^{(1+ \alpha)/2}$ scalings  we
have $x \propto ({\cal S}_\alpha)^{1/(1 +\alpha)}$. We emphasize that this
is a property of simple binomial random walks  which we can  now
exploit to investigate the solution of the QTM. 
More specifically this scaling implies 
\begin{equation}
P_B(x,{\cal S}_\alpha) = {1 \over ({\cal S}_\alpha )^{1/(1+\alpha)}}
 B_\alpha\left[{ x \over ({\cal S}_\alpha)^{1/(1 +\alpha)} } \right].   
\label{eqSca03}
\end{equation} 
Here $B_\alpha(z)$ is a a non negative function
normalized according to $\int_{-\infty} ^\infty B_\alpha(z) {\rm d} z =1$.
Further from symmetry of the walk $B_\alpha(z) = B_\alpha(-z)$. 
As  shown in Fig. \ref{fig1},
the PDF $B_\alpha(z)$ exhibits an interesting transition between a
$V$ shape for $\alpha \to 0$ and   a Gaussian shape when $\alpha \to 1$. 
In the following sections we will investigate $B_\alpha(z)$ in detail. 

\begin{figure}
\begin{center}
\epsfig{figure=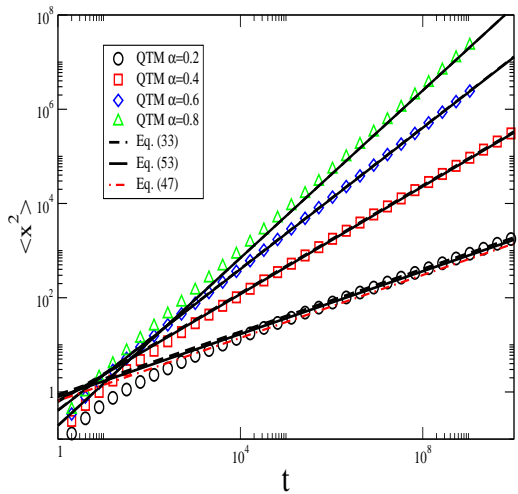,totalheight=0.34\textheight, width=0.45\textwidth,angle=-90}
\end{center}
\caption{ The mean square displacement
versus time. Numerical data obtained from direct simulations of
the QTM match  perfectly the theory based on 
weak subordination breaking (the lines plotted using
Eq. 
(\ref{eqSca08})
and Table \ref{table1}).
Analytical predictions for $\langle x^2 \rangle$ 
Eq. (\ref{eqaplz13}) for $\alpha=0.2$ and Eq.
(\ref{eqBa204}) for $\alpha=0.4,0.6,0.8$ 
perfectly match 
direct simulations and weak subordination 
scheme (based on Brownian simulations) and
hence the corresponding curves fall one on top of the other.
}
\label{fig2}
\end{figure}

 With $B_\alpha(z)$  we obtain useful relations between the
diffusion front of the trap model and Brownian motion.
Define the dimensionless time $\tilde{t} = t/ A^{1/\alpha} $ 
and the scaling variable
\begin{equation}
\xi = { x \over \left( \tilde{t} \right)^{{\alpha \over 1 + \alpha}} }.
\label{eqSca04}
\end{equation} 
Then it is easy to show that
\begin{equation}
\langle P(x,t) \rangle  \sim {  g_\alpha \left( \xi \right)\over \left( \tilde{t}\right)^{{ \alpha \over 1 + \alpha }} }
\label{eqSca05}
\end{equation} 
and using Eq. (\ref{eqP01}) 
\begin{equation}
g_\alpha\left(\xi\right) = \int_0 ^\infty {\rm d} y y^{ { \alpha \over 1 + \alpha }} B_\alpha \left( \xi y^{{ \alpha \over 1 + \alpha }} \right) l_{\alpha,1,1}\left( y \right) . 
\label{eqSca06}
\end{equation} 
For the behavior of $\langle P(x,t)\rangle$
 on the origin 
we use 
\begin{equation}
\int_0 ^\infty {\rm d} y y^{ q} l_{\alpha,1,1}\left(y\right) = \left\{
\begin{array}{l l} 
\infty & \mbox{if} \ \  q/\alpha > 1 \\
 \ & \ \\
{ \Gamma\left( 1 - q/\alpha\right) \over \Gamma\left( 1 - q \right) } & \mbox{if} \ \ q/\alpha<1
\end{array}
\right.
\label{eqEql}
\end{equation}
and then find
\begin{equation}
\langle P(x=0,t) \rangle \sim B_\alpha(0) { \Gamma\left( { \alpha \over 1 + \alpha} \right) \over
\Gamma \left( { 1 \over 1 + \alpha } \right) \left( \tilde{t} \right)^{{\alpha \over 1 + \alpha}}  } . 
\label{eqSca07}
\end{equation} 
This is a useful result since the behavior 
of $B_\alpha(z)$ on the origin $z=0$
gives the corresponding behavior of $\langle P(x=0,t) \rangle$ without
the need to solve any integral. Further Eq.
(\ref{eqSca07}) hints to an interesting behavior
when $\alpha \to 0$. The ratio of the $\Gamma$ functions
diverges in that limit,  hence as shown in Fig. \ref{fig1}
$B_\alpha(z=0)$ must go
to zero when $\alpha \to 0$ for $\langle P(x=0,t) \rangle$ to remain
finite.  Such a behavior is analytically investigated in the following
section.

 Another useful relation is found between the moments 
$\langle |x|^q \rangle = \langle 
\int_{-\infty} ^\infty |x|^q P(x,t) {\rm d} x \rangle
$ of the original QTM and the moments 
$\langle |z|^q \rangle=\int_{-\infty} ^\infty |z|^q B_{\alpha} (z) {\rm d}z$. 
Using Eqs.
(\ref{eqSca06},
 \ref{eqEql})
we find
\begin{equation}
\langle |x|^q \rangle = \langle |z|^q \rangle { \Gamma\left( { q \over 1 + \alpha} \right) \over \alpha \Gamma \left( { q \alpha \over 1 + \alpha } \right) } \left( \tilde{t} \right)^{{\alpha q \over 1 + \alpha} } .
\label{eqSca08}
\end{equation} 
The scaling $x^2 \propto 
\left( \tilde{t} \right)^{{2 \alpha  \over 1 + \alpha} }$
 was obtained long ago
in \cite{Alexander1,Bouchaud}
using elegant scaling arguments 
and in  \cite{Machta}  using renormalization group approach. 
The new content of Eqs.
(\ref{eqSca06},
\ref{eqSca07},\ref{eqSca08}) is that once we obtain $B_\alpha(z)$ 
either from theory or simulations of Brownian trajectories,
 we have a useful method to obtain
exact  statistical properties of the diffusion front.

On a computer our approach is very useful. 
For example we
have numerically generated Brownian trajectories on a lattice and
obtained $B_\alpha(z)$ in Fig. \ref{fig1} while
 $\langle z^2 \rangle$ and $B_\alpha(0)$ are reported in Table \ref{table1}. 
With $\langle z^2 \rangle$ given  in Table \ref{table1} 
and Eq. (\ref{eqSca08}) we get the mean square displacement of the
QTM $\langle x^2 \rangle$. Direct simulations
of the QTM are favorably  compared with
the predictions of  our theory in Fig. \ref{fig2}.

 Finally the cumulative distribution function $G_\alpha(\xi<\Xi) = \int_{-\infty} ^\Xi g_\alpha(\xi) {\rm d} \xi$,
 the probability that the random variable $\xi$ attains a value less then $\Xi$
is found using Eq. 
(\ref{eqSca03}) 
\begin{equation}
G_\alpha(\xi< |\Xi|)  = 1 - \int_0 ^\infty {\rm d} z B_\alpha(z) L_\alpha\left[ \left( { z \over |\Xi|} \right)^{1 + \alpha \over \alpha} \right] .
\label{eqCum}
\end{equation}
Here $L_\alpha(y)= \int_0 ^y l_{\alpha,1,1} (y) {\rm d} y$ 
is the cumulative distribution of a one sided stable random variable.
From symmetry 
$G_\alpha\left( \xi , - |\Xi|\right) = 1 - G_\alpha ( \xi < |\Xi|)$. 
The integral representation of the
distribution $L_\alpha(y)$ can be found in \cite{Chambers}. 

\begin{figure}
\begin{center}
\epsfig{figure=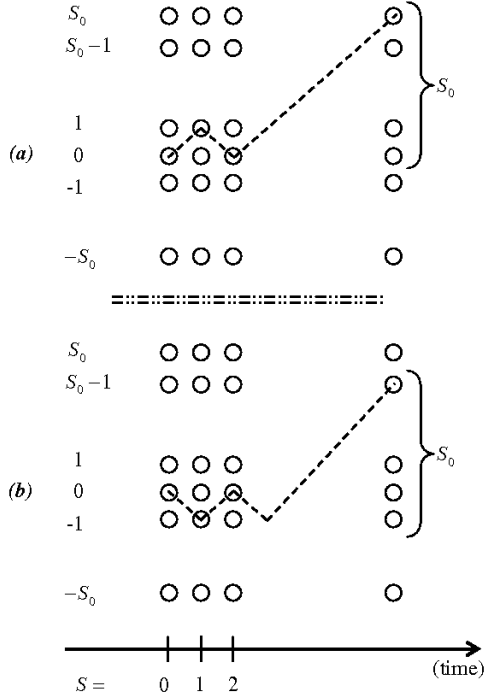,totalheight=0.34\textheight,angle=0,width=0.45\textwidth}
\end{center}
\caption{ 
(a)  Trajectory with a span ${\cal S}_0$. 
The particle returns to the origin along its path
[category (ii) in text], hence in the time interval $s>1$ (excluding
the first step) the displacement is ${\cal S}_0 - 1$ and the span ${\cal S}_0$. 
(b) A 
random walker with span ${\cal S}_0$ reaching $x={\cal S}_0-1$ 
must pass through $x=-1$. Here the first jump event 
brings the particle to $x=-1$ and hence the
span in $s>1$ is ${\cal S}_0$ and the displacement ${\cal S}_0$.  
}
\label{figtraj}
\end{figure}

\section{Limit $\alpha \to 0$}
\label{Secalz}

 As mentioned in the introduction the diffusion front $\langle P(x,t)\rangle$ 
was treated using a renormalization group method by Monthus \cite{Monthus}.
We will now investigate this interesting limit using our new approach.
For that we must first find $B_\alpha(z)$ in the limit $\alpha \to 0$.

\subsection{$\lim_{\alpha \to 0} B_\alpha(z)$ has a V shape}  

 We consider Brownian motion on a lattice which is stopped
once 
the distinct number of sites visited by the walker
 reaches the threshold  ${\cal S}_0$ and as a reminder
${\cal S}_0$ is called the span. The position of the
particle is then $x$ and we are interested in the probability
$P(x,{\cal S}_0)$ of finding the particle on $x$.   

 The particle starts on the origin, hence clearly we have $|x| \le {\cal S}_0$.
Further $P(x=0,{\cal S}_0) = 0$, since a particle starting on the origin
cannot reach the threshold ${\cal S}_0$  when it is on the origin:
i.e. a particle returning to the origin is not increasing ${\cal S}_0$ since
the origin is not a new site visited by the walker. 
From symmetry $P(-x,{\cal S}_0) = P(x,{\cal S}_0)$.

Consider first $P(x={\cal S}_0 ,{\cal S}_0)$. After the first step the
 particle
can be either on $x=1$ with probability $1/2$ 
or on $x=-1$ with the same probability. Clearly a trajectory going through
 $x=-1$ cannot contribute to   
$P(x={\cal S}_0 ,{\cal S}_0)$ since to reach $x={\cal S}_0$ through $x=-1$ the
span must be at-least of length ${\cal S}_0 + 1$. So we consider
only trajectories going through $x=1$. Trajectories going through
$x=1$ are divided into three non-intersecting
categories. Trajectories that (i)
 never reach the origin $x=0$ along their path,
 (ii) trajectories that reach the origin but never cross it
see Fig. \ref{figtraj}(a)
 and (iii) 
trajectories that go below $x=0$.
The latter will have a total span greater than ${\cal S}_0$ and hence do not
contribute.  For class $(i)$ the span (after stepping into $x=1$ in the
first step) is ${\cal S}_0 -1$. 
Similarly for class $(ii)$ the span is ${\cal S}_0$.
For both cases the displacement (from $x=1$ to ${\cal S}_0$) is clearly ${\cal S}_0 - 1$. 
Hence we have
$$ P(x= {\cal S}_0,{\cal S}_0) = $$
\begin{equation}
 { 1 \over 2} \left[  P\left( x= {\cal S}_0 - 1,{\cal S}_0 \right) +  P \left( x= {\cal S}_0 - 1 ,{\cal S}_0-1\right) \right]
\label{eqalpz01}
\end{equation}  
where the first (second) term on the right hand side describes trajectories 
returning  (never returning) to the origin. 
The half in front of the square brackets is due to the displacement 
in the first jump event. 
 
 Continuing with similar reasoning consider 
$P\left( x= {\cal S}_0 -1,{\cal S}_0\right)$. The 
particle after the first step can
be either on $x=1$ or $x=-1$. As shown in Fig. 
\ref{figtraj}(b), if it is on $x=-1$ it must travel
a distance ${\cal S}_0$ to reach its destination ${\cal S}_0 - 1$ while
keeping the  span ${\cal S}_0$. On the other hand
if it jumps to  $x=1$ 
the distance the particle must travel is ${\cal S}_0 - 2$ and
the span is ${\cal S}_0$. Hence we have
\begin{equation}
P\left( x= {\cal S}_0 - 1, {\cal S}_0 \right) = { 1 \over 2} P\left( x= {\cal S}_0, {\cal S}_0 \right) + { 1 \over 2} P\left( x = {\cal S}_0 - 2 , {\cal S}_0 \right)  
\label{eqalpz02}
\end{equation}  
where the first (second) term on the right had side describes trajectories
starting on the origin and in the first step jumping
 to  $x=-1$ $(x=1)$ respectively.  
Similarly for ${\cal S}_0 - n >0$ with $n>0$ being an integer we
have 
$$ P\left(x={\cal S}_0 -n , {\cal S}_0 \right) =$$
\begin{equation}
{ 1 \over 2} P\left(x={\cal S}_0 -n-1 , {\cal S}_0 \right) +
{ 1 \over 2} P\left(x={\cal S}_0 -n+1 , {\cal S}_0 \right). 
\label{eqalpz03}
\end{equation}  
Eqs. (\ref{eqalpz01},\ref{eqalpz03}) are easily solved
\begin{equation}
P\left(x,{\cal S}_0 \right) = { |x| \over {\cal S}_0 \left( {\cal S}_0 + 1 \right) } \ \ \mbox{for} \ \ \ -{\cal S}_0   \le x \le {\cal S}_0  
\label{eqalpz04}
\end{equation}  
and $x \in {\bf Z}$. 
% \mathbb{Z} does not work 
In the limit ${\cal S}_0\gg 1$ we have for the scaled variable
 $z=x/{\cal S}_0$
the PDF
\begin{equation} 
\lim_{\alpha \to 0} B_{\alpha} (z) = \left\{
\begin{array}{l l}
|z| & \mbox{for} |z|<1 \\
\  & \  \\
0 & \mbox{otherwise} 
\end{array}
\right.
\label{eqaplz05}
\end{equation} 
We see that $B_{\alpha = 0 } (z)$ has a $V$ shape. This reflects the
observation that the  particle reaching a large span ${\cal S}_0$ is most
likely far from the origin, and the probability of reaching the span
${\cal S}_0$ for the first time, while the particle is on the origin
being zero. We now use this property of Brownian motion to solve the quenched
trap model in the limit $\alpha \to 0$.  

\subsection{Diffusion front in the $\alpha \to 0$ limit} 

Define
the Fourier transform 
of the scaling function $g_\alpha(\xi)$ Eq.
(\ref{eqSca05})
\begin{equation} 
 g_\alpha(k_\xi) = \int_{-\infty} ^\infty e^{ i k_{\xi} \xi} g_\alpha(\xi) {\rm d} \xi  
\label{eqaplz06}
\end{equation} 
which as usual is also the moment generating function 
\begin{equation} 
 g_\alpha(k_\xi) = \sum_{q=0} ^\infty { (i k_\xi)^{2 q} \langle \xi^{2 q } \rangle  \over \left( 2 q \right)!} .  
\label{eqaplz07}
\end{equation} 
According to our theory the moments 
$\langle (\xi)^{2 q} \rangle = \int_{-\infty} ^\infty g_\alpha(\xi) \xi^{2 q } {\rm d } \xi$ 
and similarly $\langle x^{2q} \rangle$ for the QTM
are determined by Brownian motion with the help of
the PDF  $B_\alpha(z)$
Eq. (\ref{eqSca08}). 
 In the limit $\alpha \to 0$ we
find using Eq. 
(\ref{eqaplz05})
\begin{equation}
\langle z^{2 q} \rangle = 2 \int_0 ^1 z^{2 q}  z {\rm d} z = { 1 \over 1 + q} 
\label{eqaplz08}
\end{equation} 
and hence for $\alpha \to 0$
Eqs. 
(\ref{eqSca04},\ref{eqSca08})  
give
\begin{equation}
\langle \xi^{2 q} \rangle = { ( 2 q)! \over q + 1}  . 
\label{eqaplz09}
\end{equation} 
Therefore
\begin{equation} 
\lim_{\alpha \to 0}  g_\alpha(k_\xi) =  \sum_{q=0} ^\infty (-1)^q \left( { 1 \over q+1} \right) \left( k_\xi \right)^{2 q}, 
\label{eqaplz10} 
\end{equation} 
summing this series we find
\begin{equation}
\lim_{\alpha \to 0} g_\alpha(k_\xi) = { \ln \left[ 1 +  (k_\xi)^2 \right] \over \left( k_{\xi} \right)^2 } . 
\label{eqaplz11} 
\end{equation} 
Inverse Fourier transform 
yields
\begin{equation} 
\lim_{\alpha \to 0} g_\alpha(\xi) = e^{-|\xi|}-|\xi| E_1 \left( |\xi|\right) 
\label{eqaplz12} 
\end{equation} 
where $E_1(\xi) = \int_\xi ^\infty (e^{- t} / t){\rm d} t$ is the tabulated
exponential integral \cite{Abr}. This result (written in a different but
equivalent form) 
was obtained by C. Monthus  \cite{Monthus} using the
renormalization group method, which is exact
in the limit $\alpha \to 0$.
In Fig. \ref{fig3} we show $g_\alpha(\xi)$ 
for  simulations of the QTM $(\alpha=0.1)$,
Brownian simulations using weak subordination breaking outlined is
Sec. \ref{SecWSB} and analytical curve Eq. (\ref{eqaplz12}).
We see that the theory which is exact when $\alpha \to 0$ works well
also for small values of $\alpha$.

 When $\alpha$ is small we find a
useful approximation for the moments. Inserting $\langle |z|^{q} \rangle$
Eq. (\ref{eqaplz08}) in Eq. (\ref{eqSca08}) we have
\begin{equation}
\langle |x|^q \rangle \simeq { 2 \over 2 + q} { \Gamma\left( { q \over 1 + \alpha} \right) \over \alpha \Gamma\left( { \alpha q \over 1 + \alpha} \right)} \left( 
{ t \over A^{1 /\alpha} } \right)^{\alpha q \over 1 + \alpha}. 
\label{eqaplz13} 
\end{equation} 
Notice that in this limit
$\Gamma[q/(1+\alpha)]/\{\alpha\Gamma[\alpha q/(1+\alpha)]\}\simeq \Gamma(1+q)$ 
hence for $q=0$ we have $\langle |x|^0 \rangle= 1$ as expected
from normalization.
In Fig. \ref{fig2} we show $\langle x^2 \rangle$ versus time for $\alpha=0.2$.
Numerical simulation  of the QTM  perfectly
match Eq. (\ref{eqaplz13}). Note that the theory based on Table \ref{table1} 
and Eq. (\ref{eqSca08})
does a slightly better 
job since that approach is not limited to the $\alpha<<1$ regime.

\section{Approaching the Gaussian limit $\alpha =1$}
\label{Secalo}

 In this section we consider the case $\alpha \to 1$ from below.
We now find an approximation for $B_\alpha(z)$  which yields
the
solution of the QTM in this limit.

\subsection{ $B_\alpha (z) $ is Gaussian when $\alpha \to 1$} 

As before to find $B_\alpha(z)$ we consider Brownian motion. 
The probability of finding the particle on
$x$ at time $s$ is a Gaussian
\begin{equation}
P(x,s) = { \exp\left( -{ x^2 \over 2 s} \right) \over \sqrt{ 2 \pi s} }
\label{eqBa101} 
\end{equation}
as is well known. 
For $\alpha=1$ we have ${\cal S}_1 = \sum_{x=-\infty} ^\infty n_x= s$, namely
${\cal S}_1$ is not a random variable at all since it 
 is  equal to the number of steps made in the underlying random
walk. 
In other words the PDF of ${\cal S}_1$
is a delta function centered on $s$.
Therefore  when
$\alpha$ is close enough to $1$  we
may neglect fluctuations.  This means that
 we omit the
average in 
Eq. (\ref{eqApC14zz}) and use 
 ${\cal S}_\alpha = {\cal C}_\alpha s^{{1 + \alpha \over 2 }}$. 
This approach together with Eq. (\ref{eqBa101}) gives 
the PDF of finding the particle
on $x$ for a random walk stopped at the $\alpha$ coverage time
${\cal S}_\alpha$ 
\begin{equation}
P_B\left(x,{\cal S}_\alpha\right) \simeq { \exp\left[ - { x^2 \over 2 \left( {\cal S}_\alpha / {\cal C}_\alpha\right)^{{ 2 \over 1 + \alpha}} }\right] \over 
\left[ 2 \pi \left( {\cal S}_\alpha / {\cal C}_\alpha \right)^{{ 2 \over 1 + \alpha}} \right]^{1/2} }. 
\label{eqBa102} 
\end{equation} 
Hence  
\begin{equation}
B_\alpha(z) \simeq { \exp\left[ - { \left( {\cal C}_\alpha \right)^{{ 2 \over 1 + \alpha}} z^2  \over 2} \right] \over \left[ 2 \pi / ({\cal C}_\alpha)^{ { 2 \over 1 + \alpha}} \right]^{1/2} },
\label{eqBa102a} 
\end{equation} 
and it follows that 
\begin{equation} 
\langle z^2 \rangle= \left( {\cal C}_\alpha \right)^{-{ 2 \over 1 + \alpha }}.
 \label{eqBa203}
\end{equation} 
In Fig. \ref{fig1} $B_\alpha(z)$ obtained
from Brownian simulations is compared with the  analytical prediction
Eq. (\ref{eqBa102a}) for $\alpha=0.9$.  

\begin{figure}
\begin{center}
\epsfig{figure=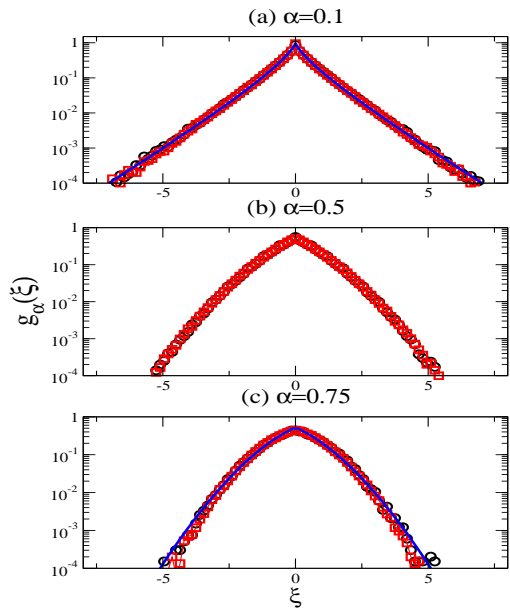,totalheight=0.34\textheight, width=0.45\textwidth}
\end{center}
\caption{ The diffusion front  of QTM (simulations) nicely
matches our theory based on weak sub-ordination breaking
[the algorithm in Sec. (\ref{SecWSB})] 
and analytical predictions Eqs. 
(\ref{eqaplz12}, 
\ref{eqBa102cc}) 
for $\alpha=0.1,0.75$ respectively.  
}
\label{fig3}
\end{figure}

\subsection{$\langle P(x,t) \rangle$ when $\alpha \simeq 1$}

From Eqs. 
(\ref{eqP01},\ref{eqBa102}) 
we have 
\begin{equation}
\langle P(x,t) \rangle \simeq \int_0 ^\infty {\rm d} {\cal S}_\alpha  { \exp\left[ - { x^2 \over 2 \left( {\cal S}_\alpha / {\cal C}_\alpha\right)^{{ 2 \over 1 + \alpha}} } \right] \over 
\left[ 2 \pi \left( {\cal S}_\alpha / {\cal C}_\alpha \right)^{{ 2 \over 1 + \alpha}} \right]^{1/2} } n\left({\cal S}_\alpha , t \right) . 
\label{eqBa102cc} 
\end{equation} 
From Eq. (\ref{eqBa102cc}) it is easy to get $g_\alpha(\xi)$ which
is given in Eq.
(\ref{eqApD01}) in Appendix D. 
In Fig. \ref{fig3} we compare between the scaling function
obtained analytically and numerical simulations of the QTM,
and with Brownian simulations according to the disorder free algorithm in Sec.
\ref{SecWSB}. The theory works reasonably well even for $\alpha = 0.75$.

Using Eqs. 
(\ref{eqSca08},\ref{eqBa203}) 
we find the mean square displacement of the
QTM
\begin{equation}
\langle x^2 \rangle \simeq \left( {\cal C}_\alpha\right)^{-{ 2 \over 1 + \alpha }} { \Gamma\left( { 2 \over 1 + \alpha} \right) \over \alpha \Gamma\left( { 2 \alpha \over 1 + \alpha } \right) } \left( { t \over A^{1/\alpha} } \right)^{{ 2 \alpha \over 1 + \alpha}} . 
\label{eqBa204}
\end{equation} 
Calculation of other moments is as simple, since the reader may easily obtain
$\langle |z|^q \rangle$ from the Gaussian PDF Eq. 
(\ref{eqBa102a}) and then apply  Eq. (\ref{eqSca08}).  
The behavior on the origin is found using Eqs.
(\ref{eqSca07},\ref{eqBa102a}) 
\begin{equation}
\langle P\left(x=0,t \right) \rangle \simeq {\left( {\cal C}_\alpha \right)^{1 \over 1 + \alpha} \over \sqrt{2 \pi} } { \Gamma\left( { \alpha \over 1 + \alpha } \right) \over \Gamma\left( { 1 \over 1 + \alpha } \right)} \left( \tilde{t} \right)^{ - {\alpha \over 1 + \alpha} } . 
\label{eqBa205}
\end{equation}
When $\alpha = 1$ we get the expected behavior 
$\langle P(x=0,t) \rangle = ( 2 \pi \tilde{t})^{-1/2}$,
which is normal diffusion.  

\begin{figure}
\begin{center}
\epsfig{figure=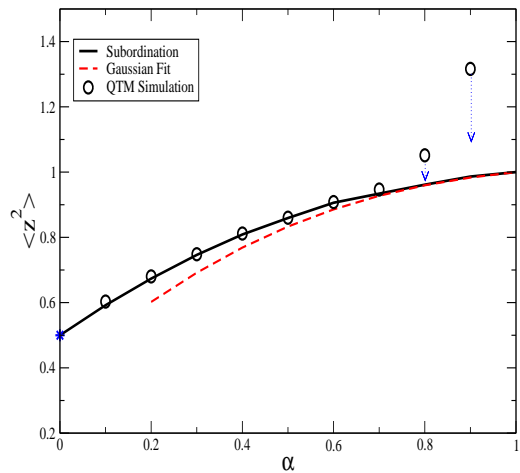,totalheight=0.34\textheight, width=0.45\textwidth,angle=-90}
\end{center}
\caption{ 
We show $\langle z^2 \rangle$ versus $\alpha$. According to theory
$\langle z^2 \rangle = 1/2$ for $\alpha \to 0$ and $\langle z^2 \rangle =1$
when $\alpha \to 1$. We compute $\langle z^2 \rangle$ using Brownian
 simulations as in Table \ref{table1}
 (solid curve) and with the QTM.
 For $\alpha=0.8$ and $\alpha = 0.9$ the finite time simulations of the QTM did not converge, as
shown in Fig. \ref{fig5}. However extrapolating the data (see Fig. \ref{fig5})
we get excellent agreement between simulations of the trap model and
our theory  based on weak sub-ordination breaking.  
}
\label{fig4}
\end{figure}

 The scaling function $g_\alpha\left( \xi \right)$ is analyzed in Appendix D.
Using properties of stable PDFs, we  show that
when $\xi <<1$ 
\begin{equation}
g_\alpha \left( \xi \right) \sim g_\alpha (0) - 2^{{\alpha -1 \over 2} } \left( { 1 + \alpha \over \alpha} \right) { {\cal C}_\alpha \over \Gamma\left( { 1 - \alpha \over 2} \right) } \xi^\alpha + \cdots 
\label{eqBa206}
\end{equation}
with
\begin{equation}
g_\alpha(0) = { \left( {\cal C}_\alpha \right)^{{ 1 \over 1 + \alpha }} \over \sqrt{ 2 
\pi} } { \Gamma \left( { \alpha \over 1 + \alpha } \right) \over \Gamma \left( { 1 \over 1 + \alpha } \right) }  .
\label{eqBa206a}
\end{equation}
In the limit $\alpha \to 1$ we use $\lim_{\alpha\to 1} {\cal C}_\alpha=1$ and
 Eq. (\ref{eqBa206}) gives
\begin{equation}
\lim_{\alpha \to 1} g_\alpha \left( \xi \right) \sim { 1 \over \sqrt{ 2 \pi} } - \lim_{\alpha \to 1} {2 \over \Gamma\left( { 1 - \alpha \over 2}\right)  } \xi^\alpha + \cdots.
\label{eqBa207}
\end{equation}
The first term clearly reflects an ordinary Gaussian diffusion front.
 The second term 
vanishes in the limit $\alpha =1$ since $\Gamma(0) = \infty$. This is because
$g_1(\xi)$ is Gaussian and hence the second term in
the expansion must be a
$\xi^2$ term. So the $1/\Gamma(0)$ kills the $\xi^\alpha$  
in Eq. (\ref{eqBa207}) as $\alpha\to 1$.  
In the opposite limit of $\xi>>1$ a steepest descent method gives 
\begin{equation}
g_\alpha \left( \xi \right) \sim b_1 \xi^{ - 2 { 1 - \alpha \over 3 - \alpha }} e^ { - b_2 \xi^{2 { 1 + \alpha \over 3 - \alpha }}} 
\label{eqBa208}
\end{equation} 
where $b_1$ and $b_2$ are found in Appendix D.
In the limit we find 
\begin{equation}
\lim_{\alpha \to 1} g_\alpha (\xi)  = { 1 \over \sqrt{ 2 \pi} } e^{ - {\xi^2 \over 2}}
\label{eqBa208a}
\end{equation} 
the expected Gaussian behavior. 

\begin{figure}
\begin{center}
\epsfig{figure=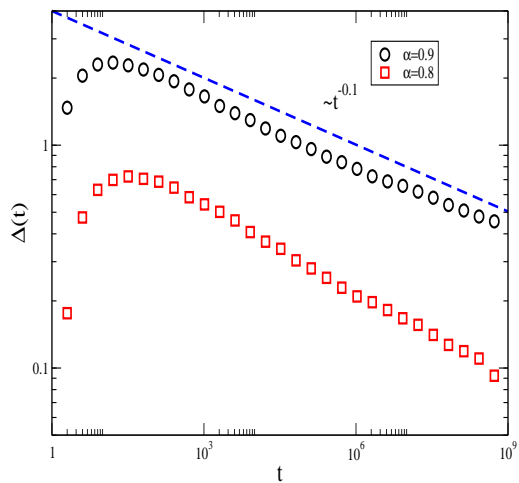,totalheight=0.34\textheight, width=0.45\textwidth,angle=-90}
\end{center}
\caption{  
 Here we show that $\Delta$ defined in Eq.
(\ref{eqDel})
 approaches 
zero extremely slowly when $\alpha=0.8$ and $\alpha=0.9$. 
Following the observation of Bertin and Bouchaud,
close to the transition from anomalous to normal
types of diffusion we find critical slowing down. 
We see that $\Delta(t) \simeq 0.1$ at times $10^9$ when $\alpha=0.8$
and hence
simulations did not converge at a time scale 
which is on the verge of our numerical capabilities 
(for $\alpha=0.9$ the situations is worse). 
By extrapolation
 the figure shows that the asymptotic
theory is reached albeit slowly. 
The dashed line is a guide to the eye with a $t^{-0.1}$ behavior. 
}
\label{fig5}
\end{figure}

\begin{figure}
\begin{center}
\epsfig{figure=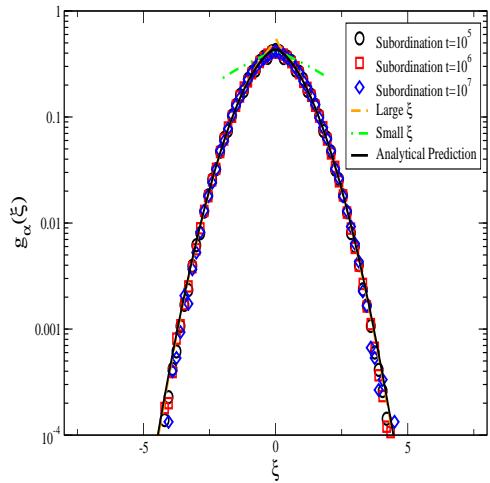,totalheight=0.34\textheight, width=0.45\textwidth,angle=-90}
\end{center}
\caption{  
For the case $\alpha=0.9$ we present the scaling PDF $g_\alpha(\xi)$ obtained
using weak subordination scheme Sec. \ref{SecWSB} 
together with  analytical curve Eq.  
(\ref{eqBa102cc}) and large and small $\xi$ expansions Eqs.  
(\ref{eqBa206},
\ref{eqBa208}).
 Similar direct simulation of the quenched
trap model do not converge even when $t=10^9$ (see Figs. \ref{fig4},\ref{fig5}) 
The figure demonstrates that with our approach 
based on weak sub-ordination breaking
reasonable converges is found  already for $t=10^5$. Hence
close to the critical point $\alpha=1$ weak subordination breaking 
can deal with the critical slow down.  
}
\label{fig7}
\end{figure}

\section{Critical slow down $\alpha \to 1$}
\label{secSim}

 As discussed in \cite{Bertin}  close to
the critical point $\alpha=1$ convergence  of
direct simulations of the QTM
is extremely slow. 
In contrast simulations of Brownian trajectories, using weak sub-ordination
scheme  
converges in reasonable time, at least on our computer. 
In this sense our approach
is much more efficient
compared with direct simulation of the QTM. 
In fact we believe that our scheme is 
the only numerical tool available today for the investigation
of the limit $\alpha \to 1$.

 In Fig.
\ref{fig4}  we show $\langle z^2 \rangle$
versus $\alpha$. $\langle z^2 \rangle$ was obtained by several means: i) 
simulation of the QTM which give $\langle x^2 \rangle$,
 and then with Eq. 
(\ref{eqSca08}) we extract $\langle z^2 \rangle$ (ii) Brownian simulations on a lattice (results in Table
\ref{table1}), and (iii) analytical theory Eqs. 
(\ref{eqaplz08},
\ref{eqBa203}). 
For $\alpha>0.8$ our simulations of the QTM did not converge even for $t =10^9$.
To check this issue better we define the deviation
\begin{equation}
\Delta(t) \equiv | {\langle x^2 \rangle \alpha \Gamma\left({2 \alpha \over 1 + \alpha} \right) \over
\Gamma\left( { 2 \over 1 + \alpha} \right) \left( t / A^{1/\alpha} \right)^{ {2 \alpha \over 1 + \alpha }}  } - \langle z^2 \rangle | .
\label{eqDel}
\end{equation} 
According to Eq. (\ref{eqSca08}) $\lim_{t \to \infty} \Delta(t) = 0$.
In Fig. \ref{fig5} we present $\Delta(t)$ versus time. Here
 $\langle x^2 \rangle$ is obtained from QTM simulations
 and $\langle z^2 \rangle$ from
Brownian trajectories (see Table \ref{table1}). 
 In Fig. \ref{fig5}  we show that
$\Delta (t) \sim t^{ \alpha -1}$ for $\alpha =0.9$
and observe  a very slow
decay towards the asymptotic value  $\Delta(t)\to 0$. 
Simulations of the QTM did not converge (even for $t=10^9$)
however extrapolating
the data (assuming nothing dramatic happens for times larger
than $10^9$) we can conclude that $\Delta(t) \to 0$ and in that sense
our theory is consistent with the simulations. 

 To over come critical slow down in the region $\alpha \to 1$ 
we use weak subordination scheme Sec. \ref{SecWSB} instead of
direct numerical simulations of the QTM. 
In Fig. \ref{fig7} we show the diffusion front.  Good agreement 
between analytical predictions (valid for $\alpha \to 1$ ) 
Eqs.  
(\ref{eqBa102cc},
\ref{eqBa206},
\ref{eqBa208})
and weak subordination breaking algorithm is presented for $\alpha =0.9$.

\section{Discussion} 
\label{SecDIS} 

  The main focus of this paper was  on  the diffusion front $\langle P(x,t) \rangle$  of random walkers
in  the quenched
trap model in one dimension. We showed that $\langle P(x,t) \rangle$ 
is found with
a L\'evy time transformation acting on
Brownian motion  stopped at the operational 
time ${\cal S}_\alpha$. Thus we map the random walk in
disordered environment to a  
Brownian motion which is stopped at the $\alpha$ coverage time
 ${\cal S}_\alpha$. 
This new type of Brownian motion is interesting on its own right.
 For example we
have found a transition from a $V$ shape to a Gaussian behavior for the
scaling function $B_\alpha(z)$ describing this motion. Properties of 
this function determine the  statistics of diffusion in the
 QTM. For $\alpha$ close to $1$ and $0$ 
we obtained
analytical expressions for $B_\alpha(z)$ and 
$\langle P(x,t)\rangle$ while numerical
information easily obtained from Brownian simulations provide a detailed
description  of the diffusion front in the range $0<\alpha<1$. 

 For $\alpha \to 0$ our formulas reduce to the
renormalization group results obtained by C. Monthus \cite{Monthus}. 
The approach presented
here is an alternative to the renormalization group method. Its advantage
 is that it is  capable
of dealing with the whole spectrum of $\alpha$, at least numerically, 
including in the critically slowed down regime of $\alpha \to 1$. 

 Is our method  general or is it limited to the one dimensional quenched
trap model?  Clearly our approach can be extended to higher dimensions,
or  for random walks with biases. As mentioned in the introduction beyond
the critical dimension, the QTM belongs to the domain
 of attraction of the CTRW.   
Hence for an ordinary random walk
on a lattice  in dimension three we expect that 
${\cal S}_\alpha$ is non-random and equal to  
  $c_\alpha s$ when time $s$ is large, and $c_\alpha$ 
is a constant so far not determined. 
In that case  usual subordination
method works for the corresponding trap model. 
Hence once
the constant $c_\alpha$ and the diffusion coefficient of
the corresponding discrete time random walk  are  determined we have 
the statistical information
needed for the determination the diffusion front of the QTM. 
Detailed analysis of the QTM for dimensions
higher than one and for biased processes, using methods developed
here, are left for future work.

$$ $$
{\bf Acknowledgement} This work was supported by the Israel science foundation.
We thank Satya Majumdar for his correspondence on the Feynman-Kac approach
which can be used to  derive Eq.  
(\ref{eqApC13}). 
Special thanks to Zvi Shemer for many discussion along this project. 

\newpage
\section{Appendix A}

 Here we summarize some known results on one sided L\'evy stable
random variables, which are used all along this work.
 By definition $l_{\alpha,1,1}(t)$ is the inverse Laplace transform
of $\exp(- u^\alpha)$. The large $t$  series expansion 
\begin{equation}
l_{\alpha,1,1} \left( t \right)={1\over \pi}\sum_{n = 1} ^\infty 
{ \Gamma\left( 1 
 + n \alpha\right) \over n!} \left( - 1 \right)^{ n - 1} \sin \left( \pi n
 \alpha \right) t^{ - (\alpha n + 1 ) } .
\label{eqA12}
\end{equation}
 The asymptotic small $t$ behavior is \cite{Schneider1}
\begin{equation}
l_{\alpha,1,1}(t) \sim B t^{-\sigma} e^{- \kappa t^{-\tau}}
\label{eqA13}
\end{equation}
where
$$ \tau={\alpha \over 1 - \alpha} , \ \kappa=(1-\alpha) \alpha^{\alpha/(1-\alpha)} , \sigma={2 -\alpha \over 2 ( 1 - \alpha)}, $$
\begin{equation} 
  B=\{ \left[2 \pi \left(1 - \alpha\right)\right]^{-1} \alpha^{1 / (1 -\alpha)} \}^{1/2}.
\label{eqA13a}
\end{equation}
Closed form PDFs are found by summing the series Eq. (\ref{eqA12})
for specific choices of $\alpha$ \cite{BarkaiPRE,Penson}. 
For example we insert the series
Eq. (\ref{eqA12}) in Mathematica and use the command Simplify to get L\'evy
PDFs in terms of a combination of Hypergeometric functions
(e.g. for $\alpha =1/4$).  Similarly L\'evy's PDFS with
 $\alpha=1/4,1/3,1/2,2/3,9/10$ can be
expressed in terms of special functions. In this way we construct 
stable distributions. Some care must be practiced
since  in some occasions we found that 
for extremely small $t$ Mathematica yields wrong results \cite{BarkaiPRE}.
This problem is easily
fixed since we can use Eq. (\ref{eqA13}) in that regime.
 Further the problem is
not  crucial in the
sense that it is found for so small $t$ that practically
the PDF there is zero, though if one is not aware of this issue solving 
integral
 transformations like Eq.
(\ref{eqBa102}) 
 can lead to wrong results.    
A useful special case is $\alpha=1/2$ 
since
\begin{equation}
l_{1/2,1,1}(t) =
{ 1 \over 2 \sqrt{\pi} } t^{ - 3/2} \exp\left( - { 1 \over 4 t} \right).
\label{eqlev} 
\end{equation}

Chambers et al \cite{Chambers} show how to generate a stable
 random variable
we call  $\eta$ from a one sided L\'evy PDF $l_{\alpha,1,1}(\eta)$. 
Let
\begin{equation}
 a(\theta) = { \sin \left( \left( 1 - \alpha \right) \theta\right) \left( \sin (\alpha \theta) \right)^{\alpha/ (1 - \alpha)} \over \left( \sin \theta\right)^{1 / (1 - \alpha) }} \ , \ 0<\theta<\pi.  
\label{eqlev01} 
\end{equation}
Then $\eta= [ a(\theta) / W ]^{(1-\alpha)/\alpha}$ where $\theta$ is a
 uniform random number
on $(0,\pi)$ and $W$ is a random variable drawn from a standard  
exponential distribution:
 $W= -\ln(x)$ were $x$ is uniform on $(0,1)$.

\section{Appendix B}

In this Appendix we  obtain the distribution of $\eta$ 
defined in Eq. (\ref{eqtime02}).
%\end{equation} 
%
We are interested in random walks 
with fixed  ${\cal S}_\alpha$  
 and in the limit ${\cal S}_\alpha \to \infty$. A large ${\cal S}_\alpha$ 
implies also a large number of steps (denoted with $s$),  however 
since ${\cal S}_\alpha$ is fixed $s$ remains random.
Our starting point is Eq. 
(\ref{eqApA02z})
\begin{equation}
\langle e^{- u \eta } \rangle = \Pi_{x=-\infty} ^\infty \hat{\psi} \left[ { n_x u \over  ({\cal S}_\alpha)^{1/\alpha} }  \right] .
\label{eqApA02}
\end{equation} 
It is important to notice that the visitation numbers $\{ n_x \}$ are determined
by the probabilities of jumping left and right (equal $1/2$ in our model) and
that these random numbers do not depend on the waiting times since
here ${\cal S}_\alpha$ is fixed. 
 Hence statistics of these
visitation numbers  are determined by simple binomial random walks.  

 The Laplace $\tau\to u$  transform of a rather 
general waiting time PDF is
in the small $u$ limit
\begin{equation}
\hat{\psi}(u) = 1 - A u^\alpha + B u ^\beta + \cdots
\label{eqApA03}
\end{equation} 
where as mentioned $0<\alpha<1$, $A>0$ and $\beta> \alpha$. The goal
is to show that  when ${\cal S}_\alpha \to \infty$ 
parameters like $B$
and $\beta$ are not important. To see this insert Eq. (\ref{eqApA03}) in
Eq. (\ref{eqApA02}) and find
\begin{equation}
\langle e^{ - u \eta} \rangle = \Pi_{x=-\infty} ^\infty \left[ 1 - A { (n_x)^\alpha \over \left( {\cal S}_\alpha \right) } u^\alpha + B {(n_x)^\beta \over ({\cal S}_\alpha)^{\beta/\alpha} } u^\beta + \cdots \right] .
\label{eqApA04}
\end{equation} 
This can be rewritten as
\begin{widetext}
\begin{equation}
 \langle e^{ - u \eta} \rangle = 
1 - A u^\alpha + \sum_{x=-\infty} ^\infty \sum_{y=-\infty,y\neq x} ^\infty 
{A^2 \over 2} {(n_x)^\alpha (n_y)^\alpha \over \left( {\cal S}_\alpha \right)^2}u^{2 \alpha} + B { \sum_{x = -\infty} ^\infty  (n_x)^\beta  
\over \left( {\cal S}_\alpha \right)^{\beta/\alpha} } u^\beta + \cdots.
\label{eqApA05}
\end{equation} 
We note that 
\begin{equation}
 \sum_{x=-\infty} ^\infty \sum_{y=-\infty,y\neq x} ^\infty (n_x)^\alpha (n_y)^\alpha=
\sum_{x=-\infty} ^\infty \sum_{y=-\infty} ^\infty 
(n_x)^\alpha (n_y)^\alpha - \sum_{x=-\infty} ^\infty (n_x)^{2 \alpha} = ({\cal S}_\alpha)^2 - {\cal S}_{2 \alpha}, 
\label{eqApA07}
\end{equation} 
hence
\begin{equation}
 \langle e^{ - u \eta} \rangle = 
1 - A u^\alpha + 
{A^2 \over 2} { ({\cal S}_\alpha)^2 - {\cal S}_{2 \alpha} \over ({\cal S}_\alpha)^2}  u^{2 \alpha} + B { {\cal S}_\beta \over ({\cal S}_\alpha)^{\beta / \alpha} } 
 u^\beta + \cdots.
\label{eqApA05a}
\end{equation} 
\end{widetext}
We use  ${\cal S}_{2 \alpha} / ({\cal S}_{\alpha})^2 \to 0$ 
which is justified at the end of this Appendix
and hence
\begin{equation}
 {({\cal S}_\alpha)^2 - {\cal S}_{2 \alpha}  \over ({\cal S}_\alpha)^2} \to 1
\label{eqApA05b}
\end{equation} 
when ${\cal S}_\alpha \to \infty$. 
Similarly
${\cal S}_\beta / ({\cal S}_\alpha)^{\beta/\alpha} \to 0$ for $\alpha < \beta$.
Summarizing  we find
\begin{equation} 
\langle e^{ - \eta u } \rangle \sim  1 - A u^\alpha + {A^2 u^{2 \alpha} \over 2} + \cdots = e^{ - A u^\alpha} 
\label{eqApA09}
\end{equation} 
The parameters $B$ and $\beta$ are unimportant. Further 
for a typical path there is no trace of the random variables $\{ n_x \}$ 
in
the final expression Eq. (\ref{eqApA09}). 
The latter Eq.  implies that the PDF of $\eta$ is a one side L\'evy PDF
as stated in Eq. 
(\ref{eqtime04}). 
 
To better estimate the convergence to this law  we use a result
obtained in Appendix C.  There we show that  
for  a binomial random walk
with $s$ steps 
we have 
\begin{equation}
\langle {\cal S}_\alpha \rangle = {\cal C}_\alpha s^{(1 + \alpha)/2} 
\label{eqApA06}
\end{equation}
where ${\cal C}_\alpha$ is a constant 
Eq. (\ref{eqApC14}). 
We then assume the following relation to hold
\begin{equation} 
{\cal S}_\alpha = r s^{(1 + \alpha)/2} 
\label{eqApA06a}
\end{equation} 
where $r$ is a random variable
which is independent of the number of steps $s$. 
Since in the QTM jumps to nearest neighbors
have probability $1/2$ like the binomial random walk and  since the statistics
of the  visitations numbers $\{ n_x \}$ are
  independent of the waiting times (for fixed ${\cal S}_\alpha$) we may
use Eq. (\ref{eqApA06}) derived for the binomial random walk
to analyze the QTM.   
We have ${\cal S}_{2 \alpha} \propto s^{(1 + 2 \alpha)/2} $ and hence
${\cal S}_{2 \alpha} \propto ({\cal S}_\alpha)^{ (1 + 2 \alpha)/ (1 + \alpha)} $
so
${\cal S}_{2 \alpha} / ({\cal S}_\alpha)^2 \propto ({\cal S}_\alpha )^{ - 1/(1 + \alpha)} $ which goes to zero in the scaling limit ${\cal S}_\alpha \to \infty$ as we stated . 
Similarly ${\cal S}_\beta /({\cal S}_\alpha)^{ \beta/\alpha} \propto s^{ (1+ \beta)/2}/ s^{\beta(1 + \alpha)/2 \alpha} = s^{ (\alpha-\beta)/(2 \alpha)} $ which approaches zero since $\alpha< \beta$.  

%Finally we note that to derive Eq. (\ref{eqApA05b}) one must show
%%
%\begin{equation} 
% \sum_{j} \sum_{i \neq j} (n_i)^\alpha (n_j)^\alpha / \sum_i (n_i)^{2 \alpha} \to \infty
%\label{eqApA06b} 
%\end{equation} 
%%
%This Eq. being obvious in the limit $\alpha \to 0$ since then
%$\sum_{i} n_i ^0$ is the span of the random walk. For general $\alpha$
%this ratio is the sum of all the non diagonal matrix elements of 
%the matrix  $(n_i)^\alpha (n_j)^\alpha$ over the sum of the diagonal 
%terms i.e. the trace. We expect the divergence in Eq. (\ref{eqApA06b}) since
%the number of of-diagonal terms far exceeds the diagonal terms
%and since within the diffusion front of a simple binomial random
%walk (i.e. $i,j < \sqrt{s}$) $(n_i)^\alpha (n_j)^\alpha$ is of the order

\section{Appendix C}

We consider a binomial random walk on a  one dimensional
lattice. Time $s$ is discrete $s=0,1,2,...$ and the particle has 
probability one half to jump to its nearest neighbors on its left or right. 
 The walk starts on the origin $x=0$. We now calculate
the average $\langle {\cal S}_\alpha \rangle$ for an $s$ step
random walk. For that aim 
we obtain  $\langle (n_x)^\alpha\rangle$  and then sum over $x$
\begin{equation} 
 \langle {\cal S}_\alpha\rangle=\sum_{x=-\infty} ^\infty\langle (n_x)^\alpha\rangle. 
\label{eqApC000}
\end{equation}
We consider this problem in the continuum limit of the model,
namely we consider Brownian motion (see details below). Thus
 our final expression
for $\langle {\cal S}_\alpha \rangle$ describes
the limit of 
large $s$. 

 Let $P_{s,x}(n_x)$ be the probability of making $n_x$ visits on 
lattice point $x$ in the time interval $(0,s)$. 
As usual in these problems the Laplace transform
\begin{equation}
\hat{P}_{u,x} (n_x) = \int_0 ^\infty e^{ - u s} P_{s,x} (n_x) {\rm d} s.
\label{eqApC01} 
\end{equation}  
is useful. Here we already started taking the continuum limit, since 
in the discrete time random walk $s$ is not a continuous variable.
We avoid a formal transition from  a discrete random walk
 to the continuum limit, to save
space and time.    

 For a random walk starting on the origin  let $\tau$ be
the first time the particle reaches lattice point $x$
and $f_x(\tau)$ its PDF.
$\tau$ is  a first passage time
for an unbiased random walk and it distribution
is well known \cite{Redner}.  
From symmetry $f_{-x}(\tau) = f_x(\tau)$. 
The number of visits on lattice point  $x$, $n_x$,
 is determined by a first passage time
from the origin to point $x$, and then by the probability to revisit
point $x$. Due to translation symmetry of the random walk the probability
of $n_x-1$
revisits to a lattice point $x$, 
in a time interval $s-\tau$ (once reaching that point at $\tau$)
 is identical to the probability of $n_x -1$ visits on the origin 
(starting on the origin) within the same time interval.  Hence
translation symmetry gives
\begin{equation}
P_{s,x}(n_x)= \int_0 ^s f_x(\tau) P_{s - \tau,0} (n_x - 1) {\rm d} \tau.
\label{eqApC00}
\end{equation} 
Using convolution theorem $\hat{P}_{u,x}(n_x)= f_x(u) \hat{P}_{u,0} (n_x-1)$.
Here $P_{s,0}(n_0)$ is the probability to visit the 
origin $n_0$ times in the time interval $(0,s)$.   

For the origin $x=0$ we have
\begin{equation}
\hat{P}_{u,x=0} (n_0) = \left\{
\begin{array}{c c}
{1 - \hat{f}_1(u) \over u} & n_0 = 0 \\ 
\hat{f}_1 (u)^{n_0} {1 - \hat{f}_1(u) \over u} & n_0 \neq 0  .
\end{array}
\right. 
\label{eqApC03}
\end{equation} 
Here $\hat{f}_1(u)$ is the Laplace transform of $f_1 (\tau)$. 
To reason for  Eq. (\ref{eqApC03}) note that if $n_0=0$
we have $P_{s,x=0}(n_0=0)= 1 - \int_0 ^s f_1(\tau) {\rm d} \tau$
which is the probability of not returning to the origin. To see this
note that after one jump the particle is either on $x=1$
or $x=-1$ and hence for $n_0$ to remain zero the particle
must not return to origin (of-course $f_1(\tau)$ is the PDF of first
passage times from $x=1$ or $x=-1$ to the origin).  
Applying the  convolution
theorem theorem of Laplace transform
to $P_{s,x=0}(n_0=0)= 1 - \int_0 ^s f_1(\tau) {\rm d} \tau$
 we get the first line in Eq.
 (\ref{eqApC03}). Note that the 
original stay on the origin;  at time $s=0$, is not counted 
so we may have $n_0 = 0$ once the particle
never returns to origin. 
The probability that $n_0=1$ is given by
\begin{equation}
P_{s,1} (n_0 = 1) = \int_0 ^s f_1(\tau) P_{s-\tau,0} (n_0 = 0) {\rm d} \tau.
\label{eqApC03a}
\end{equation} 
Again using the convolution theorem we find Eq. (\ref{eqApC03}) for $n_0 =1$.
Similarly for $n_0>1$. 
Using Eq. (\ref{eqApC00})
it is easily shown that for $x\ne 0$
\begin{equation}
\hat{P}_{u,x} (n_x) = \left\{
\begin{array}{c c}
{1 - \hat{f}_x(u) \over u} & n_x = 0 \\ 
\hat{f}_x(u) \hat{f}_1 (u)^{n_x - 1} {1 - \hat{f}_1(u) \over u} & n_x \neq 0. 
\end{array}
\right. 
\label{eqApC02}
\end{equation} 

 The Laplace transform of the first passage time PDF is 
\begin{equation}
\hat{f}_x (u) = \exp\left( - \sqrt{2} x u^{1/2}\right). 
\label{eqApC04}
\end{equation} 
The $\sqrt{2}$ comes from the fact that the diffusion
constant is equal $1/2$, since the variance of jump lengths is unity. 
In time $\tau$, $f_x(\tau)$  is the one sided L\'evy PDF with index $1/2$ 
[see  Eq. (\ref{eqlev})] and $f_x(\tau) \sim \tau^{-3/2}$ as well known \cite{Redner}. 

 We now calculate $\langle (n_0)^\alpha \rangle_{u}$: the Laplace transform
of $\langle (n_0)^\alpha \rangle_{s}$
\begin{equation} 
 \langle (n_0)^\alpha \rangle_{u}=
\int_0 ^\infty (n_0)^\alpha \hat{P}_{u,0} (n_0) {\rm d} n_0 
\label{eqApC05}
\end{equation} 
where the integration (instead of summation) implies that we are considering
the continuum limit of the random walk (i.e. Brownian motion). 
Inserting in Eq. (\ref{eqApC05}) Eqs. 
(\ref{eqApC03},\ref{eqApC04})  we find
\begin{equation}
\langle (n_0)^\alpha \rangle_u \sim \left( {1 \over \sqrt{2} } \right)^\alpha
\Gamma\left( 1 + \alpha\right)
 u^{ - 1 - {\alpha \over 2}}, 
\label{eqApC06}
\end{equation} 
which is valid for small $u$ corresponding to large $s$. 
Using Eqs. 
(\ref{eqApC02},\ref{eqApC04})
we get the $\alpha$ moment of the visitation number for lattice point $x$ 
\begin{equation} 
\langle (n_x)^\alpha \rangle_u = e^{- x \sqrt{2} u^{1/2} }\Gamma\left( 1 + \alpha\right) \left( { 1 \over \sqrt{2}} \right)^\alpha u^{ - 1 - \alpha/2} .
\label{eqApC09}
\end{equation}
Notice that 
$\langle (n_x)^\alpha \rangle_u \sim \hat{f}_x(u) \langle (n_0)^\alpha \rangle_u$ reflecting the first arrival at $x$ and the translation 
symmetry of the lattice. 
 
Denote $\langle {\cal S}_\alpha \rangle_u$ as the Laplace $s \to u$ transform
of $\langle {\cal S}_\alpha \rangle_s$. In the Brownian limit
we replace  the summation in Eq. 
(\ref{eqApC000})
with integration
\begin{equation} 
\langle {\cal S}_\alpha \rangle_u = 2 \int_0 ^\infty \langle (n_x)^\alpha \rangle_u  {\rm d} x ,
\label{eqApC10} 
\end{equation} 
and with the help of  Eq. (\ref{eqApC09})
\begin{equation}
\langle {\cal S}_\alpha \rangle_u \sim \sqrt{2}^{1-\alpha} \Gamma\left( 1 + \alpha\right) u^{ - 3/2 - \alpha/2} .
\label{eqApC11} 
\end{equation} 
Using the Laplace pair
\begin{equation}
u^{-{ 3 + \alpha \over 2} } \to {s^{{\alpha+1 \over 2}} \over \Gamma\left( { \alpha + 3 \over 2} \right)},
\label{eqApC12} 
\end{equation} 
we find with simple identities for the Gamma function \cite{Abr},
the main result of this Appendix
\begin{equation}
\langle {\cal S}_\alpha \rangle = {\cal C}_\alpha s^{{ 1 + \alpha \over 2} } 
\label{eqApC13} 
\end{equation} 
with
\begin{equation}
{\cal C}_\alpha = { 2^{{\alpha + 3 \over 2} } \Gamma\left( 1 + {\alpha \over 2}\right) \over \sqrt{\pi} \left( 1 + \alpha \right) } .
\label{eqApC14} 
\end{equation} 
For $\alpha=1$ we have ${\cal S}_1 = \sum_{x = -\infty} ^\infty n_x = s$
a result that is retrieved from Eq. (\ref{eqApC13}) since ${\cal C}_1 = 1$.
In the limit $\alpha=0$ we have ${\cal S}_0$ 
equal to the span of the random walk, namely to
 the number of distinct sites visited by the walker. As mentioned
in the main text, in this limit
we retrieve 
Eq. (\ref{eqApC14azz}) 
  found a long time ago in \cite{Erdos}.  

\section{Appendix D} 

 Here we investigate the function $g_\alpha(\xi)$ in the limit where
$\alpha<1$ is  close to unity with the Gaussian approximation for
$B_\alpha(z)$. 
Changing variables in Eq. (\ref{eqBa102}) according to 
${\cal S}_\alpha= t^\alpha \eta^{-(1 + \alpha)/2}$ we find using the
definition  
in Eq. (\ref{eqSca05})
$$ g_\alpha \left( \xi \right) = $$
\begin{equation}
{ \left( {\cal C}_\alpha \right)^{1/(1 + \alpha)} \over \sqrt{2 \pi}} { 1 + \alpha \over 2 \alpha} \int_0 ^\infty e^{ - u \eta} \eta^{1/(2 \alpha)} l_{\alpha, 1, 1} \left( \eta^{(1 + \alpha)/(2 \alpha)} \right) {\rm d} \eta 
\label{eqApD01}
\end{equation}
where 
\begin{equation}
u=\xi^2 \left( {\cal C}_\alpha \right)^{ 2/(1 +\alpha)} /2.
\label{eqApD01z} 
\end{equation}
Eq. (\ref{eqApD01}) is a Laplace transform. Inserting in
Eq. (\ref{eqApD01})  $\xi=0$, 
changing  variables according to $y=\eta^{(1 +\alpha)/(2 \alpha)} $ and using
Eq. (\ref{eqEql}) we get $g_\alpha(\xi=0)$ Eq. 
(\ref{eqBa206a}). The small $u$ limit (corresponding to small $\xi$)
of Eq. (\ref{eqApD01}) is controlled by the large $\eta$ behavior
of 
\begin{equation} 
 \eta^{1/(2 \alpha)} l_{\alpha, 1, 1} \left( \eta^{(1 + \alpha)/(2 \alpha)} \right)  \sim {\sin \pi \alpha \over \pi} \Gamma(1 + \alpha) \eta^{-1 - \alpha/2} 
\label{eqApD02}
\end{equation}
where we used the large $\eta$ expansion of stable PDFs Eq. 
(\ref{eqA12}). Using the Tauberian theorem, 
noting that $u^{\alpha/2}$ and $\eta^{-(1 + \alpha/2) } / |\Gamma(-\alpha/2)|$
are Laplace pairs, Eqs. 
(\ref{eqApD01},\ref{eqApD02}) together with some
identities of the  Gamma function
\cite{Abr} give  Eq. 
(\ref{eqBa206}). 

In the opposite limit of large $\xi$ (i.e. large $u$) we use the small $\eta$ 
behavior of 
$\eta^{1/(2\alpha)}l_{\alpha,1,1}\left(\eta^{(1+\alpha)/(2 \alpha)} \right)$
 in Eq.  
(\ref{eqApD01}). For that  aim we use the small $\eta$ behavior of one sided
stable PDFs Eq.
(\ref{eqA13}).
We get
\begin{equation}
g_\alpha (\xi) = \bar{C} \int_0 ^\infty e^{ - u \eta} \eta^{ - \gamma} e^{ - \kappa \eta^{-\delta} } {\rm d} \eta
\label{eqApD03}
\end{equation}
where $\kappa$ is defined in Appendix B Eq. (\ref{eqA13a}), 
$$ \gamma= { 3 - \alpha \over 4 ( 1 - \alpha)}, \ \ \delta= { 1 + \alpha \over 2 (1 - \alpha)} $$ 
\begin{equation}
 \bar{C} = B {({\cal C}_\alpha)^{1/(1 + \alpha} \over \sqrt{2 \pi} } {1 + \alpha \over 2 \alpha} 
\label{eqApD02aa}
\end{equation}
and $B$ is defined in  Eq. (\ref{eqA13a}). 

 We now use steepest descent method. Let 
$h(\eta) = u \eta + \kappa \eta^{-\delta}$. 
The extremum is on $\eta_e$ which is determined
us-usual from  $\partial h/ \partial \eta = 0$ so
$\eta_e = (u / \kappa \delta)^{- 1/(1+\delta)}$. Using the expansion
\begin{equation} 
h(\eta) = h(\eta_e) + {1 \over 2} \kappa\delta ( \delta + 1 ) (\eta_e)^{ - \delta - 2} \Delta^2 + \cdots 
\label{eqApD03aa}
\end{equation}
where $\Delta= \eta- \eta_e$ is small. 
We then have after extending the domain of integration 
$$ g_\alpha(\xi) \sim  $$
\begin{equation}
\bar{C} \left( \eta_e \right)^{-\gamma} e^{ - h(\eta_e)} \int_{-\infty} ^\infty
\exp\left[ - \delta(1 + \delta) \kappa (\eta_e)^{- \delta - 2} \Delta^2  /2\right]{\rm d} \Delta.
\label{eqApD04}
\end{equation}
Solving this Gaussian integral
\begin{equation}
g_\alpha ( \xi) = \tilde{\tilde{C}} u^{ - \mu} e^{ - \bar{\kappa} u^{ \delta / ( 1 + \delta)} },
\label{eqApD05}
\end{equation}
where $\tilde{\tilde{C}} = 
\bar{C} \sqrt{2 \pi}  [ \delta ( 1 + \delta) \kappa]^{-1/2}  
 (\kappa \delta)^\mu$,
 $\bar{\kappa} = (\kappa \delta)^{1/(1 + \delta)} + \kappa(\kappa \delta)^{- \delta/ ( \delta + 1) } $
and $\mu= (1-\alpha)/(3 - \alpha)$.
Changing back to the variable $\xi$ (instead of $u$) using Eq.
(\ref{eqApD01z}) we get 
Eq. (\ref{eqBa208}) 
$$ g_\alpha(\xi) = $$
\begin{equation}
\tilde{\tilde{C}} \left[ {\xi^2 \left( {\cal C}_\alpha\right)^{{2\over 1+ \alpha}} \over 2} \right]^{-\mu} \exp\left\{ - \bar{\kappa} \left[ { \left( {\cal C}_\alpha \right)^{{2 \over 1 +\alpha}} \xi^2 \over 2} \right]^{{1 + \alpha \over 3 - \alpha}} \right\}.
\label{eqApD06}
\end{equation} 
To prepare for the limit $\alpha \to 1$ we rewrite
\begin{equation} 
 \tilde{\tilde{C}} = \left({\cal C}_\alpha\right)^{{1\over 1+\alpha}} { 1 +\alpha \over 2 \alpha} { B \over \sqrt{ \delta \left(1 + \delta\right)  \kappa}} .
\label{eqApD07}
\end{equation} 
Using 
 $\lim_{\alpha \to 1} {\cal C}_\alpha =1$,
\begin{equation}
\lim_{\alpha \to 1} { B \over \sqrt{ \delta( 1 + \delta) \kappa} } = { 1 \over \sqrt{ 2 
\pi}} 
\label{eqApD08}
\end{equation}
and
$\lim_{\alpha \to 1} \bar{\kappa}=1$
we find the expected Gaussian behavior Eq. 
(\ref{eqBa208a}).
Rewriting 
\begin{equation}
g_\alpha \left( \xi \right) \sim b_1 \xi^{ - 2 { 1 - \alpha \over 3 - \alpha }}
e^ { - b_2 \xi^{2 { 1 + \alpha \over 3 - \alpha }}}
\label{eqBa208cc}
\end{equation}
when $\xi>>1$ with
$b_1=\sqrt{(1+\alpha)/[2\pi \alpha (3-\alpha)]}D,
b_2=\left[(3-2\alpha)/2\right]D^2\ \mbox{and} \
D=\left[(1+\alpha)^{1-\alpha}\alpha^\alpha\cal{C}_\alpha\right]^{1/(3-\alpha)}$.

\end{document}